\documentclass[12pt]{amsart}

\usepackage{amsmath,amssymb}
\usepackage{graphicx}

\makeatletter
\@addtoreset{equation}{section}
\makeatother

\usepackage{enumerate}

\def\ii{{\sqrt{-1}}}
\def\ee{\mathrm e}

\def\am{\mathrm {am}}
\def\al{\mathrm {al}}

\def\angle{\mathrm{angle}}
\def\Sym{\mathrm {Sym}}

\def\PSL{\mathrm {PSL}}

\def\AA{{\mathcal A}}
\def\BB{{\mathcal B}}
\def\OO{{\mathcal O}}
\def\CC{{\mathbb C}}
\def\RR{{\mathbb R}}
\def\ZZ{{\mathbb Z}}
\def\PP{{\mathbb P}}

\newtheorem{definition}{Definition}[section]

\newtheorem{theorem}{Theorem}[section]
\newtheorem{proposition}{Proposition}[section]
\newtheorem{corollary}{Corollary}[section]
\newtheorem{remark}{Remark}[section]
\newtheorem{lemma}{Lemma}[section]

\def\book#1{\rm{#1}, }
\def\paper#1{\textit{#1}, }
\def\jour#1{\rm{#1}, }
\def\yr#1{({\rm{#1}) }}
\def\vol#1{\textbf{#1}}
\def\pages#1{\rm{#1}}
\def\page#1{\rm{#1}}

\def\publaddr#1{\rm{#1}, }
\def\publ#1{\rm{#1}, }
\def\by#1{{\rm{#1}, }}

\begin{document}

\title{Reality Conditions of Loop Solitons Genus $g$:
Hyperelliptic am Functions}
\date{}

\author{Shigeki MATSUTANI}

\maketitle

\begin{abstract}
This article is devoted to an investigation
of a reality condition of a hyperelliptic
loop soliton of
higher genus.
In the investigation, we have a natural
extension of Jacobi am-function for an
elliptic curves to that for a hyperelliptic
curve. We also compute winding numbers
of loop solitons.
\end{abstract}

\bigskip

{\centerline{\textbf{2000 MSC: 37K20, 35Q53, 14H45, 14H70 }}}

\bigskip
\section{ Introduction}

In this article, we will investigate a reality
condition of loop solitons
with genus $g$ more precisely.
Here the loop soliton is defined as follows.
\begin{definition}
A loop soliton,
$Z_{t_2} : \RR \hookrightarrow \CC$ with $t_2\in \RR$,
$t_1 \mapsto Z = X^1 + \ii X^2$
with $\partial_{t_1} Z = \ee^{\ii \phi(t_1, t_2)}$,
 is characterized by a solution of MKdV equation
\begin{gather}
   \partial_{t_2} \phi
 + \frac{1}{4}
 (\partial_{t_1} \phi)^3
 +\partial_{t_1}^3 \phi=0.
\label{eq:MKdV}
\end{gather}
\end{definition}

In \cite{Mu},
Mumford gave natural results on the moduli of
loop solitons of genus one as elasticas
in terms of $\theta$
functions, or the geometry of
the Abelian varieties of genus one.
 However in the higher genus case,
there appears a problem that
 the moduli of the Abelian varieties differs
from the moduli of Jacobian varieties.
On the investigation of loop soliton with
higher genus,
we have chosen the strategy that we use only the
data of curves themselves to avoid the
problem of excess parameters, and give some explicit
results \cite{Ma3}. We will go on to follow the strategy
to investigate the reality condition.
In this article,
 we will interpret the results of Mumford
in terms of the language of the curve in the
case of genus one and extend the scheme to
higher genus.
Section two is devoted to the reinterpretation
of Mumford results.
Section three gives the moduli of the
loop solitons of genus two, which can be easily
generalized to higher genus cases as in \S 4.

In \cite{Ma3},
we have hyperelliptic solutions of the loop soliton
as follows.
For a hyperelliptic curve $C_g$ given by
an affine equation,
\begin{gather}
\split
C_g:\quad y^2 &=  x^{2g+1}+ \lambda_{2g} x^{2g}
         +\lambda_{2g-1} x^{2g-1}+\cdots
       +\lambda_2 x^2 +\lambda_1 x  +\lambda_0 \\
     &=(x-e_1)(x-e_2) (x-e_3)\cdots(x-e_{2g})(x-e_{2g+1}),
       \label{eq:curve-g}
 \endsplit
\end{gather}
where each $e_a$ is a complex number $\CC$,
we have  coordinates of complex space $J^\infty_g :=\CC^g$
as maps from symmetric product $\Sym^g(C_g)$ to $J^\infty_g$:
\begin{gather}
         u_{g-1} =\sum_{i=1}^g u_{g-1}^{(i)},
         \quad
          u_{g} = \sum_{i=1}^g u_{g}^{(i)}, \label{eq:J_g}
\end{gather}
\begin{gather}
             u_{g-1}^{(i)} = \int^{(x^{(i)}, y^{(i)})}_\infty
               \frac{ x^{g-2}d x }{2 y},
         \quad
         u_g^{(i)} =  \int^{(x^{(i)}, y^{(i)})}_\infty
               \frac{x^{g-1} d x }{2 y}.
\label{eq:ugi}
\end{gather}

\begin{proposition}\label{prop:loopg}
A hyperelliptic solution of the loop soliton of genus $g$
 is give by
\begin{gather}
	\partial_{t_1} Z^{(a)} = \prod_{i=1}^g (x^{(i)} - e_a),
\label{eq:loop-g}
\end{gather}
where $t_1= R u_g$ and $t_2 =R( u_{g-1} - (\lambda_{2g} + e_a)^{-1} u_g)$,
for a constant positive number $R$
the curve  (\ref{eq:curve-g}) and integrals contours
which satisfy the reality condition,
\begin{enumerate}
\item $|\partial_{u_g} Z^{(a)}| = R$,

\item $u_g \in \RR$.
\end{enumerate}
\end{proposition}

\begin{proof} \cite{Ma3} Proposition 3.4. \end{proof}

In this article, we focus on the reality condition.
As shown in Theorem \ref{theorem:g}, the reality condition
is reduced to the following conditions.
\begin{theorem}\label{theorem:g}
Let a set of the
 zero points $e_b$ of $y$ in (\ref{eq:curve-g}) be denoted
by $\BB$. $Z^{(a)}$ satisfies the reality condition
if the following conditions satisfy,

\begin{enumerate}
\item each $e_c \in \BB$ is real,

\item there exists $g$ pairs $(e_{c_j}, e_{d_j})_{j=1, \cdots, g}$
satisfies $(e_{c_j}-e_a)(e_{d_j}-e_a)=e_a^2$ for
negative $e_a$,

\item the contour in the integral $u_g$
in (\ref{eq:J_g}) satisfies a condition.
\end{enumerate}
\end{theorem}

In the investigation, we have a natural
extension of Jacobi am-function for an
elliptic curves to that for a hyperelliptic
curve. We also compute winding numbers
of loop soliton.

As there are so many open problems related to
this as in \cite{Ma7, P}, this result could be
applied to them.

\bigskip

I thank
Prof. E.~Previato,  Prof. J. McKay and Prof. Y. \^Onishi
 for helpful suggestions and
encouragements. Especially I am grateful to Prof. J. McKay
for directing me to the book of Prasolov and Solovyev
\cite{PS}.

\bigskip
\section{Genus One}

First we will consider genus one case using the
data of curve given by
\begin{gather}
\split
 y^2 &=x^3
       +\lambda_2 x^2 +\lambda_1 x  +\lambda_0 \\
     &=(x-e_1)(x-e_2) (x-e_3).
 \endsplit
\label{eq:curve-g1}
\end{gather}
The coordinate $u$ of the complex plane $J_1^\infty:=\CC$
is given by,
\begin{gather}
        \int^{(x,y)} d u, \quad d u = \frac{ d x }{2 y}.
         \label{eq:u-g1}
\end{gather}

It is known that a shape of
the (classical) elastica, {\it i.e.},
 a loop soliton with genus one, $Z : \RR \hookrightarrow \CC$
$(u \mapsto Z(u) = X^1(u) + \ii X^2(u))$
with $\partial_u Z = \ee^{\ii \phi}$
satisfies the differential equation,
\begin{gather}
   a\partial_{u} (\phi)
 + \frac{1}{3}(\partial_{u}\phi)^3 +\partial_{u}^3 \phi=0,
 \label{eq:SMKdV}
\end{gather}
where $\partial_u := d /du$.

\begin{proposition}{\rm (Euler \cite{E})}
A solution of (\ref{eq:SMKdV}) is given by
$$
	\partial_u Z^{(a)} = (x - e_a),
$$
for an elliptic curve given by the form
(\ref{eq:curve-g1}), which satisfies
the  reality condition:
\begin{enumerate}
\item $|\partial_u Z^{(a)}| = 1$.

\item $u \in \RR$.
\end{enumerate}
\end{proposition}
\begin{proof} for example Proposition 3.4 in \cite{Ma3}. \end{proof}

\begin{proposition}{\rm(Mumford)}\label{Mum1}
The moduli $\Lambda$
of elastica or loop soliton of genus one is given by
the following subspece in the upper half plane
$\RR+ \ii \RR_{>0}$ modulo $\PSL(2,\ZZ)$,
$$
\Lambda := \ii \RR_{>0} \cup \left(\frac{1}{2} + \ii \RR_{>0}\right)
\quad\mathrm{modulo}\quad \PSL(2,\ZZ).
$$
\end{proposition}

The purpose of this section is to  give its proof using
only the data of the curve itself.

\begin{lemma}
For different numbers $a$, $b$ and $c$ of $\{1, 2, 3\}$,
let $\ee^{2\ii\varphi_a} :=(x-e_a)/c_{cba}$,
$e_{ab} := e_a - e_b$ and $c_{cba}:=\sqrt{e_{ca}e_{ba}}$.
The  elliptic differential of the first kind (\ref{eq:u-g1})
is
$$
	du = \frac{ d \varphi_a}
	 {\sqrt{(\sqrt{e_{ba}}-\sqrt{e_{ca}})^2
            +4 \sqrt{e_{ba}e_{ca}} \sin^2 \varphi_a}}.
$$
\end{lemma}

\begin{proof}
Direct computations give
\begin{gather*}
 \split
d x &= 2 c_{c b a} \ii \ee^{2 \ii \varphi_a} d \varphi,\\
   y &=c_{c b a} \ii \ee^{2 \ii \varphi_a}
         \sqrt{ e_{ba}(\ee^{-2\ii\varphi_a}- c_{c b a} e_{ba}^{-1})
                    (\ee^{2\ii\varphi_a}- c_{c b a}^{-1} e_{c a})}\\
    &=c_{c b a} \ii \ee^{2 \ii \varphi_a}
         \sqrt{ e_{ba}(\ee^{-2\ii\varphi_a}- \sqrt{e_{ca}/ e_{ba}})
                    (\ee^{2\ii\varphi_a}-\sqrt{e_{ba}/ e_{ca}})}\\
    &= c_{c b a}\ii \ee^{2 \ii \varphi_a}
         \sqrt{ e_{ba}+e_{c a}
 - 2 \sqrt{e_{ba}e_{c a}} \cos 2\varphi_a}.
\endsplit
\end{gather*}
The addition formula $\cos(2\varphi) = 1 - 2 \sin^2\varphi$
leads the result.
\end{proof}

Let us use the standard representations,
$$
k :=\frac{2\ii \root4\of{e_{ba}e_{ca}}}
{\sqrt{e_{ba}}-\sqrt{e_{ca}}}
$$
and then
\begin{gather}
	d u = \frac{ d \varphi_a}
	 {(\sqrt{e_{ba}}-\sqrt{e_{c a}})
           \sqrt{1-k^2 \sin^2 \varphi_a}}.
      \label{eq:am-g1}
\end{gather}
By letting $w:=\sin(\varphi_a)$, (\ref{eq:am-g1}) becomes,
\begin{gather}
\split
	d u &= \frac{d w}
 {\sqrt{(1-w^2)((\sqrt{e_{ba}}-\sqrt{e_{c a}})^2
            +4 \sqrt{e_{ba}e_{c a}} w^2)}}\\
     &=\frac{d w}
	 {(\sqrt{e_{ba}}-\sqrt{e_{c a}})\sqrt{(1-w^2)(1 - k^2 w^2)}}
\endsplit
\label{eq:w-g1}
\end{gather}

\begin{remark}\label{remark:am1}{\rm
\begin{enumerate}
\item Due to the (\ref{eq:am-g1}), we have the following
elliptic integral $u(\varphi_a)$
$$
	u(\varphi_a) = \int^{\varphi_a}_{0} \frac{d \varphi}{H_a^{[1]}(\varphi)},
$$
and its inverse function $\varphi_a(u)$  gives
$$
	\exp(\ii\varphi_a(u) ) = \sqrt{x-e_a}.
$$
As $\sqrt{(e_3-e_1)/(x-e_3)}$ is sn-function, $\varphi_a(u)$
is essentially the same as Jacobi-am
function $\am(u)$ \cite{PS}, though we need Landen-transformation.

\item Behind (\ref{eq:w-g1}), there is a kinematic system
with an energy
$$
       E = \dot w^2 + (1-w^2)(1 - k^2 w^2).
$$
\end{enumerate}
}
\end{remark}
\bigskip

For any $\varphi_a$ in a certain region $[\varphi_l, \varphi_u]$,
the reality condition of
the loop soliton $Z^{(a)}$ assumes that
the denominator in (\ref{eq:w-g1}) should be real and thus
that $\sqrt{e_{ba}e_{ca}}$ and $(\sqrt{e_{ba}}-\sqrt{e_{ca}})^2$
should be real.
$$
\Im \sqrt{e_{ba}}= \Im\sqrt{e_{ca}},\quad
\angle(e_{ba})= -\angle(e_{ca}),
$$
where $\angle(a)=\Im\log(a)$ for $a\in\CC$.
Accordingly using
the expression $e_{ba}=\beta_{ba}\ee^{\sqrt{-1}\alpha_{ba}}$,
$\alpha_{ba}\in [0,\pi)$ and
$\beta_{ba} \in \RR$,  the reality condition
 of
the loop soliton $Z^{(a)}$
require alternative cases:
\begin{enumerate}
\item $\alpha_{ba}$ and $\alpha_{ca}$vanish,
{\it i.e.}, $e_{ba}$ and $e_{ca}$ belong to $\RR$, or
\item
$\alpha_{ba}=-\alpha_{ca}$ and $\beta_{ba}=\beta_{ca}$.
\end{enumerate}
However the second case means that
$(\sqrt{e_{ba}}-\sqrt{e_{ca}})^2$ vanishes and
corresponds to $k=\infty$. Thus though it is not
important, it is interesting that the second case
can be reduced to the first case,
{\it i.e.}, $\alpha_{ca}=0$,
by transforming $\varphi_{a}$ to $\varphi_{a}-\alpha_{ca}$
due to the formula in the proof in Lemma 2.1.
Here we defined $\beta_{ba} \in \RR$
rather than $\beta_{ba} \in \RR_{\ge 0}$
due to the domain of $\alpha_{ba}$.

\begin{lemma}
The reality condition  of
the loop soliton $Z^{(a)}$
is reduced to two alternative cases:
\begin{enumerate}
\item[I-1] $e_{ba}>0$ and $e_{ca}>0$,
{\it i.e.}, $k\in \ii \RR_{\ge0}$,
$w\equiv\sin \phi_a \in [-1, 1]$.

\item[I-2] $e_{ba}\le0$ and $e_{ca}\le0$, {\it i.e.},
$k>1$ and $w\equiv\sin \phi_a \in [1/k, 1]$ or
$w\equiv\sin \phi_a \in [-1, -1/k]$.

\end{enumerate}
\end{lemma}

\begin{proof}
For general $\varphi_a\in\RR$, $u$ must be real. Hence the
candidates of $e_{ba}$'s are followings:
(I-0) $e_{ba}<0$ and $e_{ca}>0$,
(I-1) $e_{ba}>0$ and $e_{ca}>0$,
and (I-2) $e_{ba}\le0$ and $e_{ca}\le0$.

In (I-0) case
$(\sqrt{e_{ba}}-\sqrt{e_{ca}})$ has a non-trivial
angle in the complex plane, which
cannot be canceled by other terms.
(I-1) is very natural. On the case (I-2),
since the region of $\sin \phi_a$ must be a subset
of $[-1, 1]$. Noting that prefactor
$1/(\sqrt{e_{ba}}-\sqrt{e_{ca}})$ generates
the factor $\ii$,
we conclude that $k>1$ and
$\sin \phi_a \in [1/k, 1]$ or
$\sin \phi_a \in [-1, -1/k]$.
\end{proof}

\begin{figure}
\begin{center}
\includegraphics[scale=0.8]{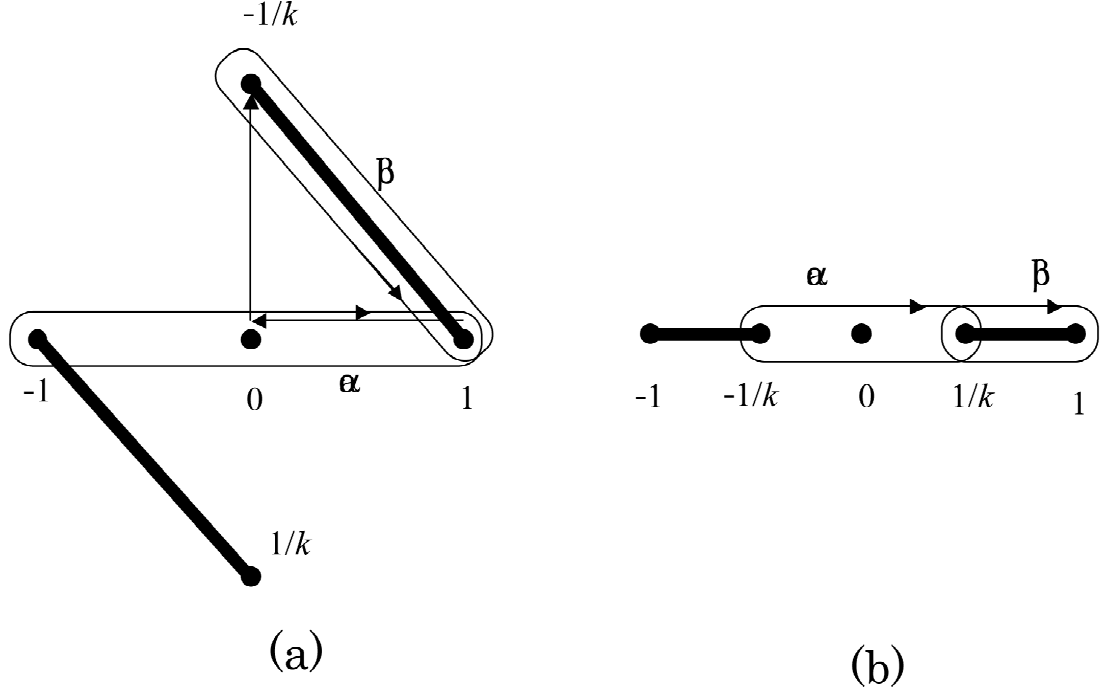}
\caption{Geometry of Contours: $\alpha$ and $\beta$ are
Homology basis of the elliptic curves.}
\label{fig:1}
\end{center}
\end{figure}
\noindent

\bigskip

{\bf Proof of Proposition \ref{Mum1}:}
Let us consider the geometry of the integration.
Fig.1 gives an illustration of our situations,
where Fig.1 (a) corresponds to case I-1 and
(b) does to case I-2.

\begin{enumerate}
\item[I-1:]
The periodicity $(4\omega, 2\omega')$ of
$\sqrt{(x-e_a)}$ is given by
\begin{gather*}
\split
	\omega &=  \int^1_0
\frac{d w}
 {\sqrt{(1-w^2)((\sqrt{e_{ba}}-\sqrt{e_{c a}})^2
            +4 \sqrt{e_{ba}e_{c a}} w^2)}},\\
	\omega' &= ( \int^0_1+\int_0^{\ii/|k|})
\frac{d w}
 {\sqrt{(1-w^2)((\sqrt{e_{ba}}-\sqrt{e_{c a}})^2
            +4 \sqrt{e_{ba}e_{c a}} w^2)}}.
\endsplit
\end{gather*}
Thus $\omega_{2} = \omega_1/2 + \ii L[k]$ for
general $k$ with a certain function $L$.
Hence $\tau = \omega'/\omega \in (1/2+ \ii \RR)$.

\item[I-2:]
The periodicity $(4\omega, 2\omega')$ of
$\sqrt{(x-e_a)}$ is given by
\begin{gather*}
\split
	\omega &= 2 \int^{1/k}_0
\frac{d w}
 {\sqrt{(1-w^2)((\sqrt{e_{ba}}-\sqrt{e_{c a}})^2
            +4 \sqrt{e_{ba}e_{c a}} w^2)}},\\
	\omega' &= \int^1_{1/k}
\frac{d w}
 {\sqrt{(1-w^2)((\sqrt{e_{ba}}-\sqrt{e_{c a}})^2
            +4 \sqrt{e_{ba}e_{c a}} w^2)}}.
\endsplit
\end{gather*}
Hence $\tau = \omega'/\omega \in \ii \RR$.

Since theory of the Jacobi elliptic functions
gives the fact that $k' :=\sqrt{1-k^2}$ gives
the inversion of moduli $\tau \to -1/\tau$,
the constraint $k>1$ is less important.
\end{enumerate}

We note that the periodicity of $\sqrt{(x-e_a)}$
differs from $\partial_u Z^{(a)}$ by twice but
the difference is not so significant.
Hence we have a complete proof of Proposition
\ref{Mum1} based upon the theory of curve
instead of geometry of Jacobain as
a domain of theta function. \qed

\begin{remark}{\rm
\begin{enumerate}
\item
We list its spacial cases for $a=1$:
\begin{enumerate}
\item $k=0$ in I-1: its shape is a circle
and its related curve is $y^2 = (x-e_1)^2(x-e_2)$

\item $k=\infty$ in I-2: its shape is a
loop soliton solution,
and its related curve is $y^2 = (x-e_1)(x-e_2)^2$
\end{enumerate}

\item
Since $\partial_s Z\equiv \ee^{\ii \phi}$
can be regarded as a harmonic
map: $\partial_s Z: S^1 \to S^1$ with energy
$$
	E = \oint d s |\partial_s \phi|^2.
$$

\item Above Lemma 2.2, we argued the angle of $e_{ba}$'s.
However the geometry of the integrals depends only
$\sqrt{e_{ba}e_{ca}}$ and $\sqrt{e_{ba}}-\sqrt{e_{ca}}$
rather than $e_{ba}$'s themselves.
\end{enumerate}}
\end{remark}

For the map $\partial_u Z: S^1 \to S^1$,
  we can find index as a winding number
as shown in Fig.2.  We call it index($\partial_u Z$).

\begin{figure}
\begin{center}
\includegraphics[scale=0.8]{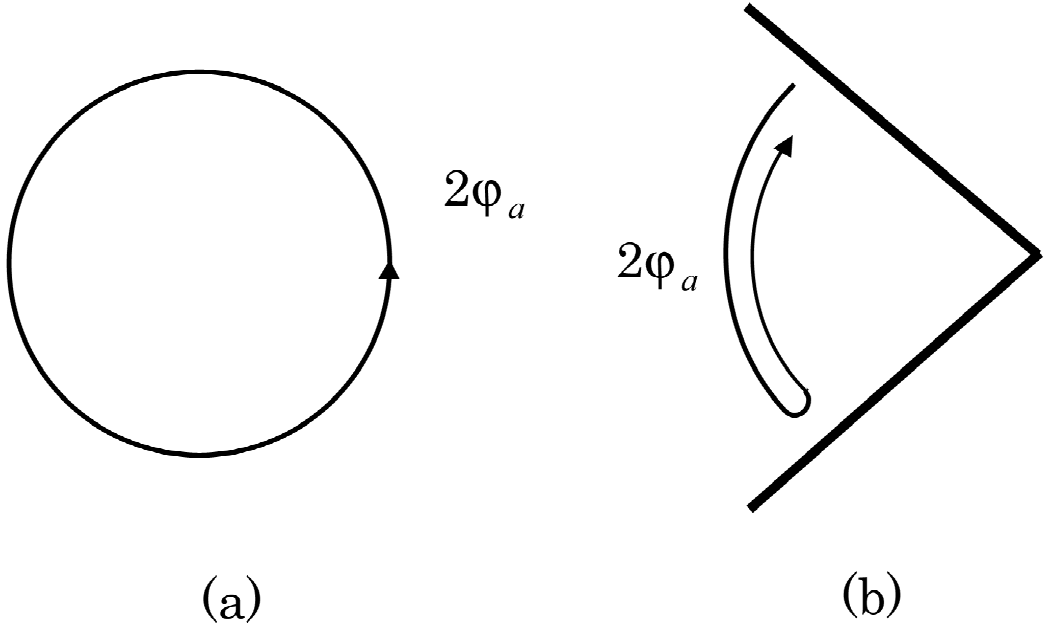}
\caption{The behavior of $\varphi$}
\label{fig:2}
\end{center}
\end{figure}
\noindent

\begin{corollary}
The $\mathrm{index}(\partial_u Z)$ is given as follows.
\begin{enumerate}
\item[I-1] $\mathrm{index}(\partial_u Z)=1$.

\item[I-2] $\mathrm{index}(\partial_u Z)=0$.
\end{enumerate}
\end{corollary}

\begin{proof}
In the case I-1,
since the contours $w\equiv \sin \varphi_a$ is $[-1,1]$
which is identified with the range of sine function,
$\varphi_a$ becomes a monotonic
 increasing function of $u$.
 In fact passing by $w=\pm1$ changes the
 sign of $\sqrt{1-w^2}$ or $\cos\varphi_a$.
On the other hand, in the case I-2,
$\varphi$ does not wind around $S^1$ like Fig 2.(b).
The branch point $(1/k, 0)$ does not have an effect
of the sign of $\sqrt{1-w^2}$.
and thus it does not
\end{proof}

\bigskip
\section{Genus Two}

In this section, we will investigate the reality conditions
of genus two.
A hyperelliptic curve of genus two is expressed by
\begin{gather}
\split
 y^2 &=  x^5 + \lambda_4 x^4 +\lambda_3 x^3
       +\lambda_2 x^2 +\lambda_1 x  +\lambda_0 \\
     &=(x-e_1)(x-e_2) (x-e_3)(x-e_4)(x-e_5),
       \label{eq:curve-g2}
 \endsplit
\end{gather}
we have a coordinate of complex space $J^\infty_2 :=\CC^2$;
\begin{gather}
         u_1 =u_1^{(1)}+u_1^{(1)},
         \quad
          u_2 = u_2^{(1)}+u_2^{(1)}, \label{eq:J_2}
\end{gather}
\begin{gather}
             u_1^{(i)} = \int^{(x^{(i)}, y^{(i)})}_\infty
               \frac{ d x }{2 y},
         \quad
         u_2^{(i)} =  \int^{(x^{(i)}, y^{(i)})}_\infty
               \frac{x d x }{2 y}.
\end{gather}

The loop soliton solution
of (\ref{eq:curve-g2}) is given by
$\partial_{t_1}Z^{(a)}=(x^{(1)}-e_a)(x^{(2)}-e_a)$.

\begin{lemma} \label{lemma:g2gene}
For different numbers $a$, $b$ and $c$ of $\{1, 2, 3\}$,
let $\ee^{2\ii\varphi_a^{(i)}} :=(x^{(i)}-e_a)/c_{cba}$,
$e_{ab} := e_a - e_b$ and $c_{cba}:=\sqrt{e_{ba}e_{ca}}$.
In general, the following relation holds:
$$
	du_2^{(i)} =
 \frac{ \ii(c_{cba}\ee^{\ii\varphi_a^{(i)}}
              +e_a\ee^{-\ii\varphi_a^{(i)}})
                 d \varphi_a^{(i)}}
	 {\sqrt{((\sqrt{e_{ba}}-\sqrt{e_{ca}})^2
            +4 \sqrt{e_{ba}e_{ca}} \sin^2 \varphi_a^{(i)})
        c_{cba}e_{da}(\ee^{-2\ii\varphi_a^{(i)}} -c_{cba}e_{da}^{-1})
                   (\ee^{2\ii\varphi_a^{(i)}} -c_{cba}^{-1}e_{ea})}}.
$$
\end{lemma}

\begin{proof}
Direct computations lead the formula.
\end{proof}

\begin{lemma} \label{lemma:RC02}
The reality condition  of
the loop soliton $Z^{(a)}$
satisfies if and only if
$(x^{(1)}, x^{(2)}) \in \Sym^2(C_2)$ and $\lambda$'s
satisfy the following relations:
\begin{enumerate}
\item $|(x^{(i)}-e_a)| = C_a$ of a real constant $C_a$, $(a=1,2)$,

\item $u_2^{(i)} \in \RR$ for $i=1,2$.
\end{enumerate}
\end{lemma}

\begin{proof}
Since the satisfaction is trivial, we will
consider the necessary condition.
$(x^{(1)}, x^{(2)}) \in \Sym^2(C_2)$ satisfying
the reality conditions becomes
$$
	|x^{(2)}-e_a| = \frac{C}{|x^{(1)}-e_a|},
$$
and
\begin{gather}
	\Im u_2^{(2)}(x^{(2)}) = - \Im u_2^{(1)}(x^{(1)}).
       \label{eq:RC-05}
\end{gather}
Assume that $|x^{(1)}-e_a|$ or $\Im u_2^{(1)}(x^{(1)})$ is
not a constant function of $x^{(1)}$. Then
$x^{(2)}$ is a function of $x^{(1)}$.
(Of course, there is no guarantee whether there exists
such a function $x^{(2)}(x^{(1)})$ and
even continuity.)
On the other hand, the loop soliton $\partial_{t_1} Z^{(a)}$
is a function of the complex space $J^\infty_2$ in
(\ref{eq:J_2})
and satisfies the MKdV equation (\ref{eq:MKdV}) over there
as mentioned in Proposition \ref{prop:loopg}.
However the assumption means that $u_1$ and $u_2$ are
not independent, {\it e.g.},
$u_2$ becomes a function of $u_1$.
It implies that $\partial_{t_1} Z^{(a)}$ a
function of a universal curve $C_2$ embedded in $J^\infty_2$
rather than $J^\infty_2$.
${\partial}/{\partial x^{(1)}}|_{x^{(2)}}$ nor
${\partial}/{\partial u_1}|_{u_2}$ do not behave well
and  should be replaced to covariant
derivatives, {\it e.g.},
${\partial}/{\partial u_1} - A_{u_1}(u_1)$,
using an appropriate connection $A_{u_1}(u_1)$.
Hence in general the assumption
 requires that the
angle part of $\partial_{t_1} Z^{(a)}$ does not satisfy
the MKdV equation (\ref{eq:MKdV}).
It is a contradiction.
\end{proof}
\bigskip
\begin{remark}{\rm
 By letting the embedding
$\iota: S^1\hookrightarrow C_2$,
$\partial_{u_2} Z$ is a analytic map from
$\Sym^2(\iota(S^1))$ to $S^1$.
}\end{remark}

\bigskip

\begin{lemma}
For the situation of Lemma \ref{lemma:g2gene},
the reality condition  of
the loop soliton $Z^{(a)}$
needs
 $e_a = - c_{cba}$ and $c_{eda}=c_{cba}$, and then
we have
\begin{gather}
	d u_2^{(i)} = \frac{ 2\sqrt{c_{c b a}}\sin\varphi_a^{(i)}
                 d \varphi_a^{(i)}}
	 {\sqrt{((\sqrt{e_{ba}}-\sqrt{e_{c a}})^2
            +4 \sqrt{e_{ba}e_{c a}} \sin^2 \varphi_a^{(i)})
           ((\sqrt{e_{d a}}-\sqrt{e_{e a}})^2
            +4 \sqrt{e_{d a}e_{e a}} \sin^2 \varphi_a^{(i)})}}.
\end{gather}
\end{lemma}

\begin{proof} Due to the Lemma \ref{lemma:RC02},
$\varphi_a^{(i)}$ is real and each factor must be
real. Hence the imaginary parts should be
canceled locally. It means the conditions.
\end{proof}

\bigskip
Let us introduce a representation as an extension
of  the standard representation (\ref{eq:am-g1}),
$$
k_1 := \frac{2\ii\root4\of{e_{ba}e_{ca}}}
{\sqrt{e_{ba}}-\sqrt{e_{ca}}}, \quad
k_2 := \frac{2\ii\root4\of{e_{da}e_{ea}}}
{\sqrt{e_{da}}-\sqrt{e_{ea}}},
$$
and then
\begin{gather}
	d u_2^{(i)}= \frac{2\root4\of{e_{ba}e_{c a}}
           \sin\varphi_a^{(i)} d \varphi_a^{(i)}}
	 {(\sqrt{e_{ba}}-\sqrt{e_{c a}})(\sqrt{e_{d a}}-\sqrt{e_{e a}})
           \sqrt{1-k_1^2 \sin^2 \varphi_a^{(i)}}
           \sqrt{1-k_2^2 \sin^2 \varphi_a^{(i)}}}.
\label{eq:am-g2}
\end{gather}
By letting $w:=\sin(\varphi_a^{(i)})$, we have
\begin{gather}
\split
	d u_2^{(i)}&= \frac{ \root4\of{e_{ba}e_{c a}}w d w}
 {\sqrt{(1-w^2)((\sqrt{e_{ba}}-\sqrt{e_{c a}})^2
            +4 \sqrt{e_{ba}e_{c a}} w^2)
((\sqrt{e_{d a}}-\sqrt{e_{e a}})^2
            +4 \sqrt{e_{d a}e_{e a}} w^2)}}\nonumber\\
     &=\frac{2\root4\of{e_{ba}e_{c a}}w d w}
	 {(\sqrt{e_{ba}}-\sqrt{e_{c a}})
           (\sqrt{e_{d a}}-\sqrt{e_{e a}})
          \sqrt{(1-w^2)(1 - k_1^2 w^2)(1 - k_2^2 w^2)}
}.
\endsplit \label{eq:am-w}
\end{gather}

\begin{remark}{\rm
\begin{enumerate}

\item (\ref{eq:am-w}) is an elliptic integral
by $u=w^2$ due to a speciality of genus two.
It cannot be generalized to higher genus case.

\item
Due to the remark \ref{remark:am1},
we should be regard that (\ref{eq:am-g2}) gives the
integral as a function $u_2^{(i)}$ of $\varphi_a^{(i)}$,
$$
     u_2^{(i)} = \int^{\varphi_a^{(i)}}_0
        \frac{d\varphi_a^{(i)\prime}}
           {H_a^{[2]}(\varphi_a^{(i)\prime})}
$$
for an appropriate function $H_2^{[2]}$.
Hence the inverse function $\varphi_a^{(i)}(u_2^{(i)})$
gives the relation,
$$
	\exp(\ii \varphi_a^{(i)}(u_2^{(i)}))
         = \sqrt{(x^{(i)}-e_a)/c_{cba}}.
$$
Further $\varphi_a:=\varphi_a^{(1)}(u_2^{(1)})
          + \varphi_a^{(2)}(u_2^{(2)})$
gives the al-function of $u_2 :=u_2^{(1)}+u_2^{(2)}$ \cite{Ba2, W1},
$$
	\exp(\ii \varphi_a(u_2)) = \al_a(u_2).
$$
Accordingly, we should regard this $\varphi_a$ as a
hyperelliptic am-function of genus two.

\item Behind the hyperelliptic
am-functions, there is also kinematic system
with a hamitonian:
$$
E= \dot w^2+
(1-w^2)((\sqrt{e_{ba}}-\sqrt{e_{c a}})^2
            +4 \sqrt{e_{ba}e_{c a}} w^2)
((\sqrt{e_{d a}}-\sqrt{e_{e a}})^2
            +4 \sqrt{e_{d a}e_{e a}} w^2).
$$

\end{enumerate}}
\end{remark}

\bigskip

For any $\varphi_a^{(i)}$ in a region $[\varphi_l, \varphi_u]$,
the reality condition of
the loop soliton $Z^{(a)}$ assumes that
the denominator should be real and thus
that $\sqrt{e_{ba}e_{ca}}$  and
$\sqrt{e_{ba}}-\sqrt{e_{ca}}$ should be also real.

\begin{theorem}
The reality condition of
the loop soliton $Z^{(a)}$ is reduced to the conditions:
$e_a = - c_{cba}$ and $c_{eda}=c_{cba}$ with
 three alternative cases:
\begin{enumerate}
\item[II-1.] $e_{ba}>0$, $e_{ca}>0$
$e_{ea}>0$, $e_{da}>0$, {\it i.e.},
$k_1, k_2\in \ii \RR$
and
$\sin \phi_a \in [-1, 1]$.

\item[II-2.]
 $e_{ba}>0$, $e_{ca}>0$,
$e_{ea}\le0$ and $e_{da}\le0$, {\it i.e.},
$k_1 \in \ii \RR $ and $k_2\in \RR$
$\sin \phi_a \in [1/k_2, 1]$ or
$\sin \phi_a \in [-1, -1/k_2]$.

\item[II-3.] $e_{ba}\le0$, $e_{ca}\le0$
$e_{ea}\le0$, $e_{da}\le0$, {\it i.e.},
 $k_1, k_2\in  \RR$, $(k_1<k_2)$,
\begin{enumerate}
\item
if $k_2<1$, $\sin \phi_a \in [-1, 1]$.

\item
if $k_2>1$, $\sin \phi_a \in [-1/k_2, 1/k_2]$.

\item
if $k_1>1$, $\sin \phi_a \in [1/k_1, 1]$ or
$\sin \phi_a \in [-1, -1/k_1]$.
\end{enumerate}
\end{enumerate}
\end{theorem}

\begin{proof}
As shown in case of the elliptic curves, we can
find the results.
\end{proof}

\begin{figure}
\begin{center}
\includegraphics[scale=0.8]{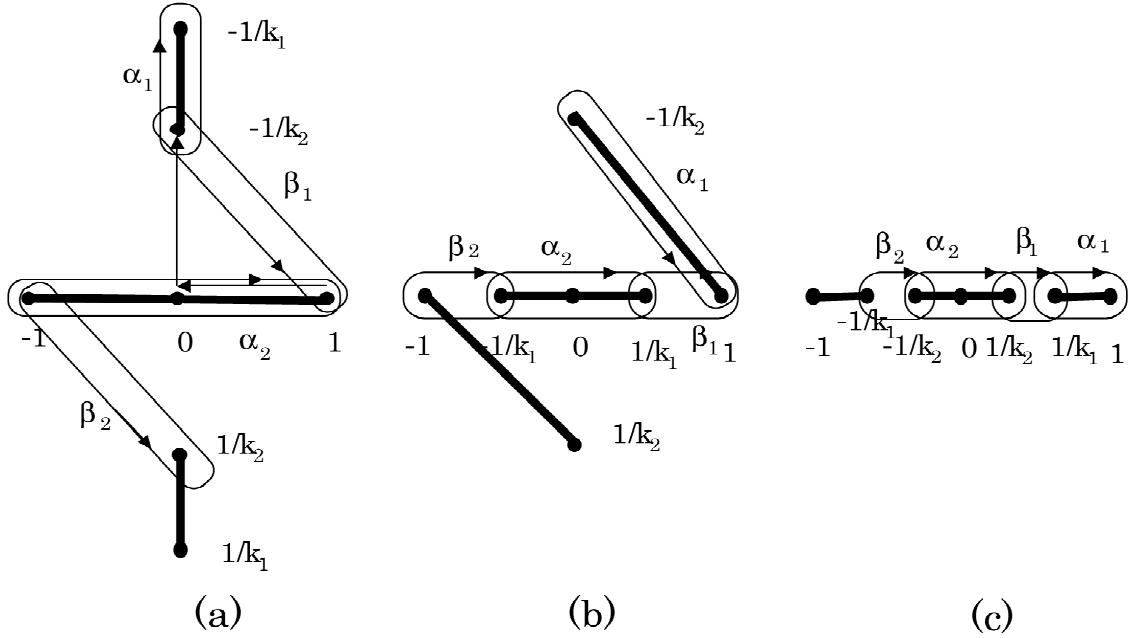}
\caption{Geometry of Contours: $\alpha_1$, $\beta_1$,
$\alpha_2$ and $\beta_2$ are Homology basis of the
hyperelliptic curves.}
\label{fig:3}
\end{center}
\end{figure}
\noindent

Fig.3 gives an illustration of our situations,
where Fig.3 (a) corresponds to II-1 and
(b) does to II-2 and (c) to II-3.

In this case,
 we show the index($\partial_{t_1} Z$).

\begin{corollary}
The $\mathrm{index}(\partial_{t_1}Z)$ as a winding
number of the map
$\iota(S^1)$ to $S^1$ is
\begin{enumerate}
\item[II-1.] $\mathrm{index}(\partial_{t_1}Z)=2$,

\item[II-2.] $\mathrm{index}(\partial_{t_1}Z)=0$,

\item[II-3.] (a) $\mathrm{index}(\partial_{t_1}Z)=2$
 and (b) (c) $\mathrm{index}(\partial_{t_1}Z)=0$.
\end{enumerate}
\end{corollary}
\begin{proof}
These indexes consist of those of each $2\varphi_{a}^{(i)}$.
If the index of $2\varphi_{a}^{(i)}$ is one,
that of $2\varphi_{a}$ is sum over $i=1,2$,
$\varphi_a = \varphi_{a}^{(1)}+\varphi_{a}^{(2)}$.
The computations of $\varphi_{a}$
are essentially the same as the genus
one illustrated in Fig.2.
\end{proof}

\bigskip
\section{Genus $g$}
The computations of genus two are easily extended to
higher genus loop solitons.
Let $\AA:=\{1, 2, 3, \cdots, 2g+1\}$,
$\AA_a:=\AA -\{a\}$ for $a\in \AA$,
$\OO_g:=\{1, 3, 5, \cdots, 2g-1\}$.
and a bijection $\sigma: \{1, 2, \cdots, 2g\} \to
\AA_a$ for $a\in \AA$, which determines
the order.

The direct computations give the following lemmas.
\begin{lemma} \label{lemma:g-gene}
For $a\in \AA$,
let $\ee^{2\ii\varphi_a^{(i)}} :=(x^{(i)}-e_a)/c_{cba}$,
$e_{ba} := e_{\sigma(b)} - e_a$ and
$c_{cba}:=\sqrt{e_{ba}e_{ca}}$,
\begin{gather}
\split
D^{(i)}_{a,\sigma}(\varphi_a):=&
\Bigr((\sqrt{e_{1a}}-\sqrt{e_{2a}})^2
            +4 \sqrt{e_{1a}e_{2a}} \sin^2 \varphi_a^{(i)})\\
            &
        \prod_{d\in \OO_g, e=d+1}
        c_{12a}e_{da}(\ee^{-2\ii\varphi_a^{(i)}} -c_{12a}e_{da}^{-1})
                   (\ee^{2\ii\varphi_a^{(i)}} -c_{12a}^{-1}e_{ea})
                    \Bigr)^{1/2},
                    \endsplit
\end{gather}
$$
N^{(i)}_{a,\sigma}(\varphi_a):=\left(
 \ii(c_{12a}\ee^{\ii\varphi_a^{(i)}}
              +e_a\ee^{-\ii\varphi_a^{(i)}})\right)^{g-1}.
$$
In general, (\ref{eq:ugi}) becomes
$$
	du_g^{(i)} =
 \frac{ N^{(i)}_{a,\sigma}d\varphi_a}{D^{(i)}_{a,\sigma}}.
$$
\end{lemma}

\begin{lemma}
For the situation of Lemma \ref{lemma:g-gene},
the reality condition of
the loop soliton $Z^{(a)}$ needs
 $e_a = - c_{cba}$ for $c \in \AA_a^o$, $b=c+1$ and then
we have
\begin{gather}
\split
D^{(i)}_{a,\sigma}(\varphi_a)&=
\Bigr(((\sqrt{e_{1a}}-\sqrt{e_{2a}})^2
            +4 \sqrt{e_{1a}e_{2a}} \sin^2 \varphi_a^{(i)})\\
 &\prod_{d\in \OO_g, e=d+1}
        ((\sqrt{e_{d a}}-\sqrt{e_{e a}})^2
            +4 \sqrt{e_{d a}e_{e a}} \sin^2 \varphi_a^{(i)})
            \Bigr)^{1/2},
             \endsplit
\end{gather}
$$
N^{(i)}_{a,\sigma}(\varphi_a)=
       \left( 2\sqrt{c_{1 2 a}}\sin\varphi_a^{(i)}\right)^{g-1}.
$$
\end{lemma}

Let $\varphi_a:=\varphi_a^{(1)}+\varphi_a^{(2)}+\cdots +\varphi_a^{(g)}$
and then  (\ref{eq:loop-g}) is expressed by
$$
	\partial_{t_1} Z^{(a)}=\ee^{2\ii\varphi_a},
$$
as a function of $u_g:=u_g^{(1)}+u_g^{(2)}+\cdots+u_g^{(g)}$.
The hyperelliptic al-function is written by
$$
	\al_a(u)=\ee^{\ii\varphi_a(u)},
$$
Hence  $\varphi_a$ can be regarded as hyperelliptic
am-function of genus $g$.

\begin{theorem}\label{theorem:g}
The reality condition of the loop soliton
$Z^{(a)}$ in (\ref{eq:loop-g})
is the conditions that  there
are $g$ pairs
$(e_{b,a},e_{b+1,a})_{b\in \OO_g}
\in \RR^2$ satisfying
$-e_a = \sqrt{e_{b,a}e_{b+1,a}}\ge 0$,
and the contour of integral
of each $u^{(i)}_g$ of $i=1, \cdots, g$ should be
chosen so that $u^{(i)}_g$ is real.
\end{theorem}

\bigskip

\bigskip

{Shigeki Matsutani}

{e-mail:RXB01142\@nifty.com}

{8-21-1 Higashi-Linkan}

{Sagamihara 228-0811 Japan}

\end{document}